\documentclass{edp-conf}
\usepackage{graphicx}
\usepackage{epsfig}
\usepackage{amssymb}
\newcommand{\gm}{\gamma}

\newcommand{\gev}{\mathrm{GeV}}
\newcommand{\tev}{\mathrm{TeV}}

\newcommand{\unitphi}{\mathrm{cm}^{-2}\:\mathrm{s}^{-1}}

\begin{document}

\TitreGlobal{SF2A 2001}
\runningtitle{F. Piron : TeV blazars as seen by the CAT telescope} 

%%-----------------------------
%%      the top matter
%%-----------------------------
\title{TeV blazars as seen by the CAT telescope} 
\author{F. Piron}\address{Groupe d'Astroparticules de Montpellier, CC 085 - B\^at. 11, Universit\'e de Montpellier II,
Place Eug\`ene Bataillon, 34095 Montpellier Cedex 5, France -- \tt{piron@in2p3.fr}}
\author{the CAT collaboration}
\maketitle
\begin{abstract}

To date, only two extragalactic objects have been firmly established as very high-energy gamma-ray sources in the Northern sky:
these are the two blazars Markarian~501 and Markarian~421.
This paper reviews the most stri\-king results obtained from these sources by the CAT atmospheric Cherenkov imaging telescope,
with a particular emphasis on the 1999-2000 and 2000-2001 observation campaigns of Markarian~421.

\end{abstract}
%
%%-----------------------------
%%      your text
%%-----------------------------
\section{Introduction}
Blazars are believed to be radio-loud active galactic nuclei (AGN) which have a relativistic jet with its axis
closely aligned to the
line of sight. Due to Doppler boosting, the jet emission dominates that of the accretion disk (and of the host galaxy) over a
large energy domain. It can extend up to the very high-energy (VHE) range ($\gtrsim$100$\gev$) in some cases.
$\tev$ blazar observations offer the possibility of investigating the physics of jets more deeply,
including particle accele\-ration and energy extraction in the vicinity of the AGN's central ``engine''.
The identification of the VHE $\gm$-ray production region with the base of the jet is indeed suggested by their
highly variable emission, which has been already observed down to sub-hour time-scales~[\cite{Gaidos96}], thus
indicating a small spatial extension via a simple causality argument.
\begin{figure}[t!]
\hbox{
{\bf(a)}\hspace*{-.4cm}\epsfig{file=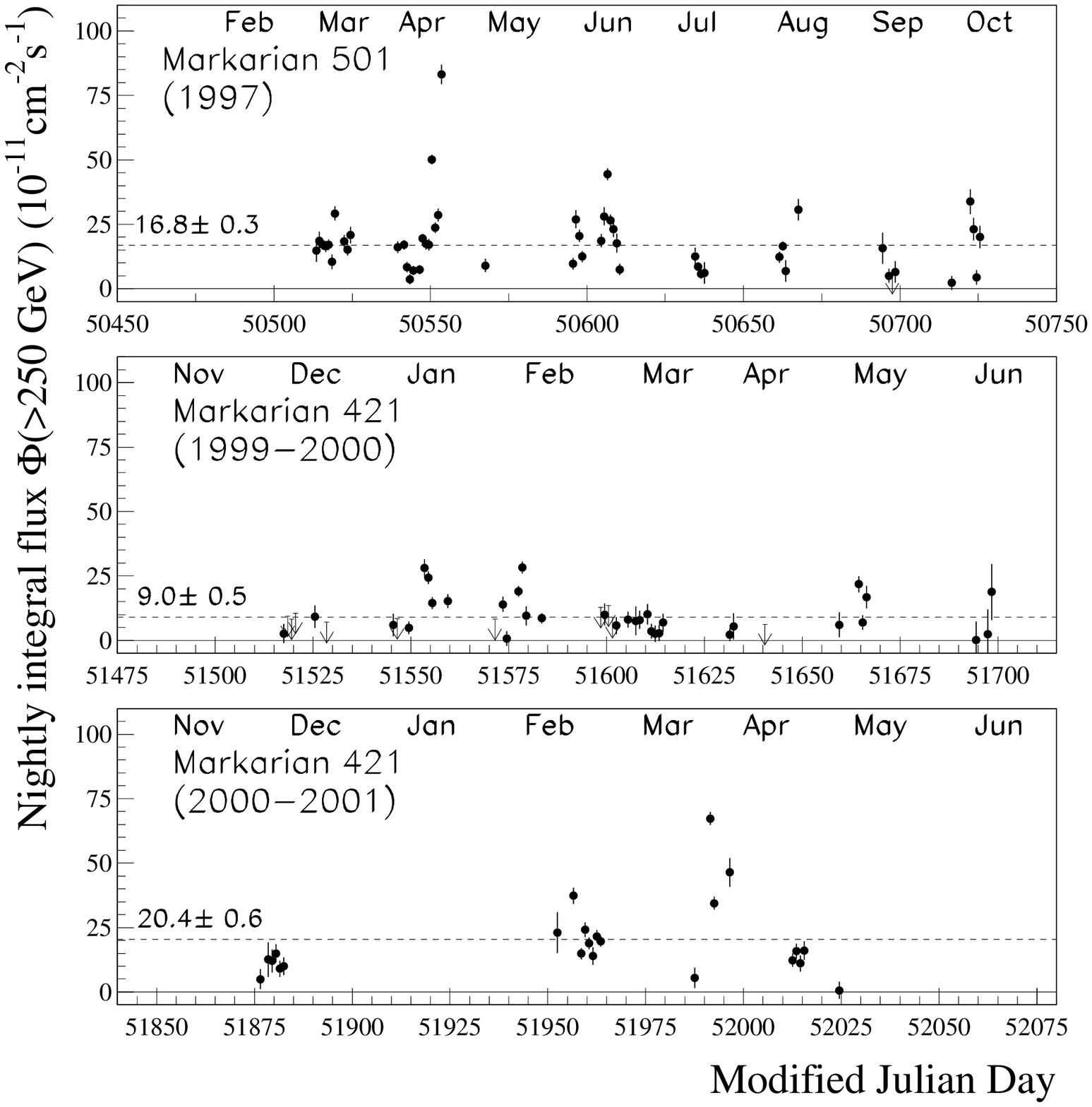,width=0.5\linewidth,clip=
,bbllx=20pt,bblly=0pt,bburx=590pt,bbury=575pt}
\hspace*{-.2cm}
\vbox{
\begin{tabular}{l}
{\bf(b)}\hspace*{-.4cm}\epsfig{file=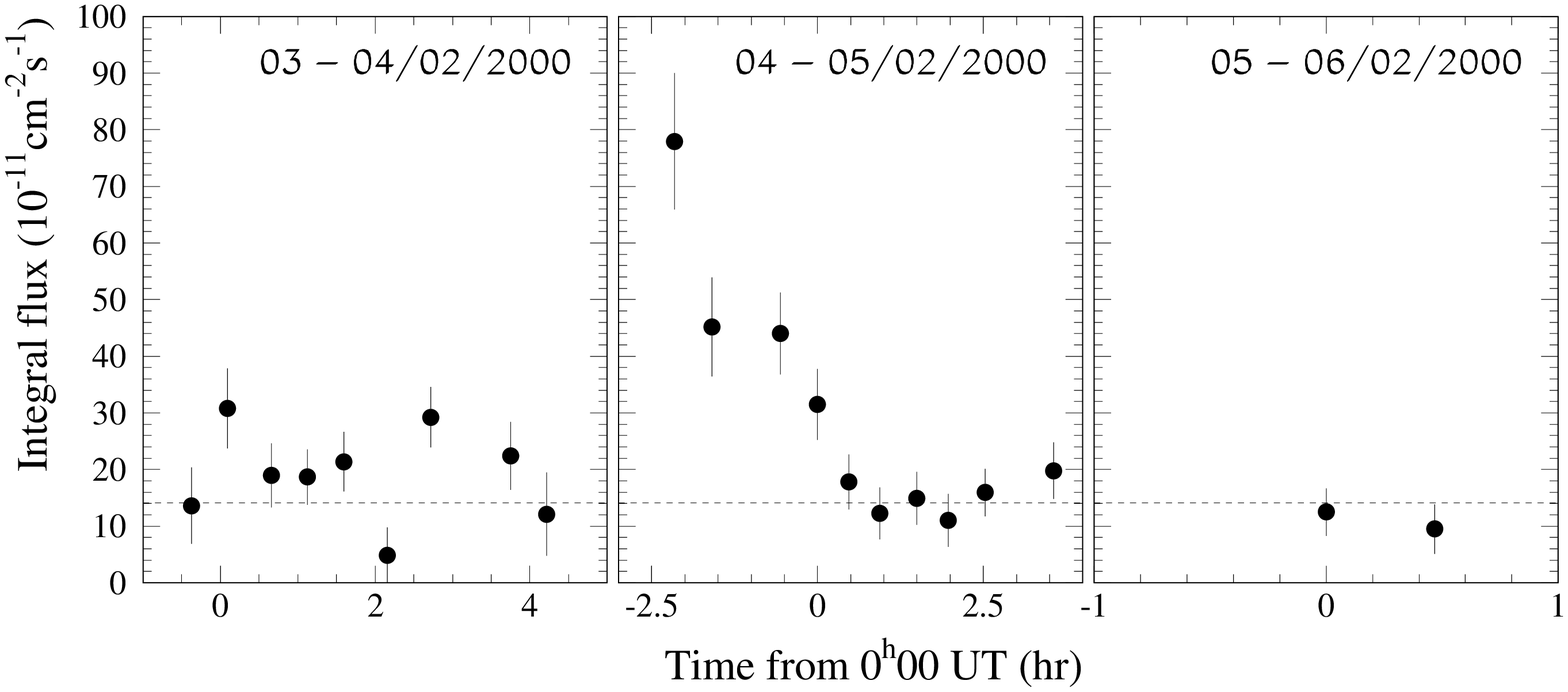,width=0.4\linewidth,clip=
,bbllx=10pt,bblly=30pt,bburx=720pt,bbury=380pt}\\\\
\vspace*{5.5cm}
{\bf(c)}\hspace*{-.4cm}\epsfig{file=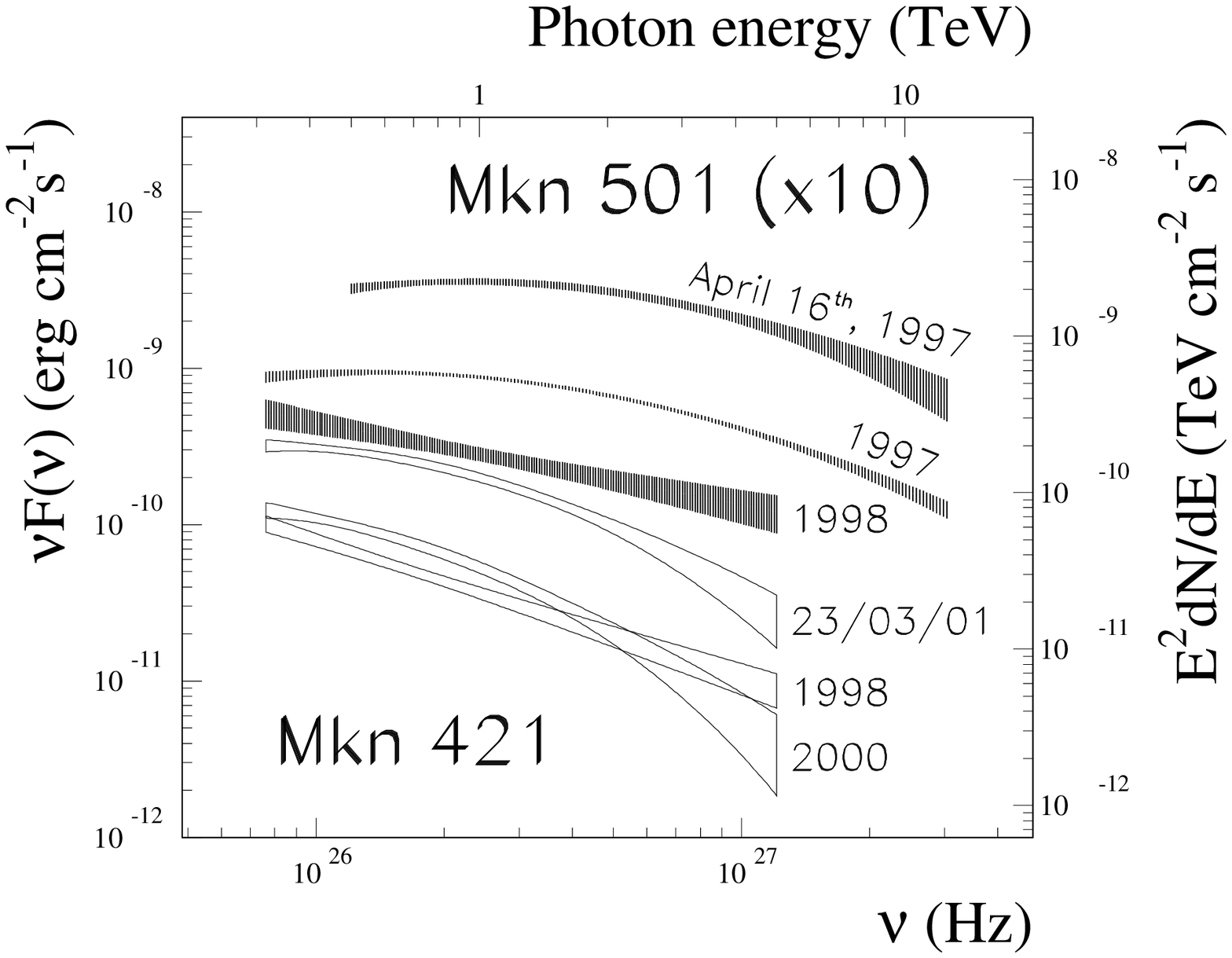,width=0.45\linewidth,clip=
,bbllx=15pt,bblly=50pt,bburx=555pt,bbury=440pt}\\
\end{tabular}
}
}
\vspace*{-5.7cm}
\caption[]
{
{\bf(a)} Mkn~501 and Mkn~421 nightly-averaged integral flux above $250\:\gev$ as recorded by CAT during their periods of most
intense activity. Dashed lines show the mean flux for each observation period;
{\bf(b)} Zoom on the nights between 3 and 6 February, 2000. Dashed lines show the Crab nebula flux;
{\bf(c)} VHE spectral energy distributions (SEDs) of Mkn~501 (hatched areas, multiplied by $10$ for clarity) and Mkn~421 (unfilled
areas). Each area stands for the $68$\% confidence level contour (see~[\cite{Piron01}] for details).
}
\label{Fig1}
\end{figure}

Since the detection by the Whipple Observatory of Markarian~421 (Mkn~421) [\cite{Punch92}] and Markarian~501
(Mkn~501)~[\cite{Quinn96}]
as the first extragalactic $\tev$ sources, blazars have been intensively observed by ground-based Cherenkov atmospheric
$\gm$-ray detectors. The CAT (Cherenkov Array at Th\'emis) imaging telescope~[\cite{Barrau98}],
which operates above $250\:\gev$ since Autumn 1996, has studied Mkn~421 and Mkn~501 in different states of activity.
The complete analyses are detailed in the literature~[\cite{Djannati99,Piron01,Khelifi01}]; here we present the main results
of this work. 

\section{VHE properties of Mkn~501 and Mkn~421}
\begin{figure}[t]
\hbox{
{\bf(a)}\hspace*{-.6cm}\epsfig{file=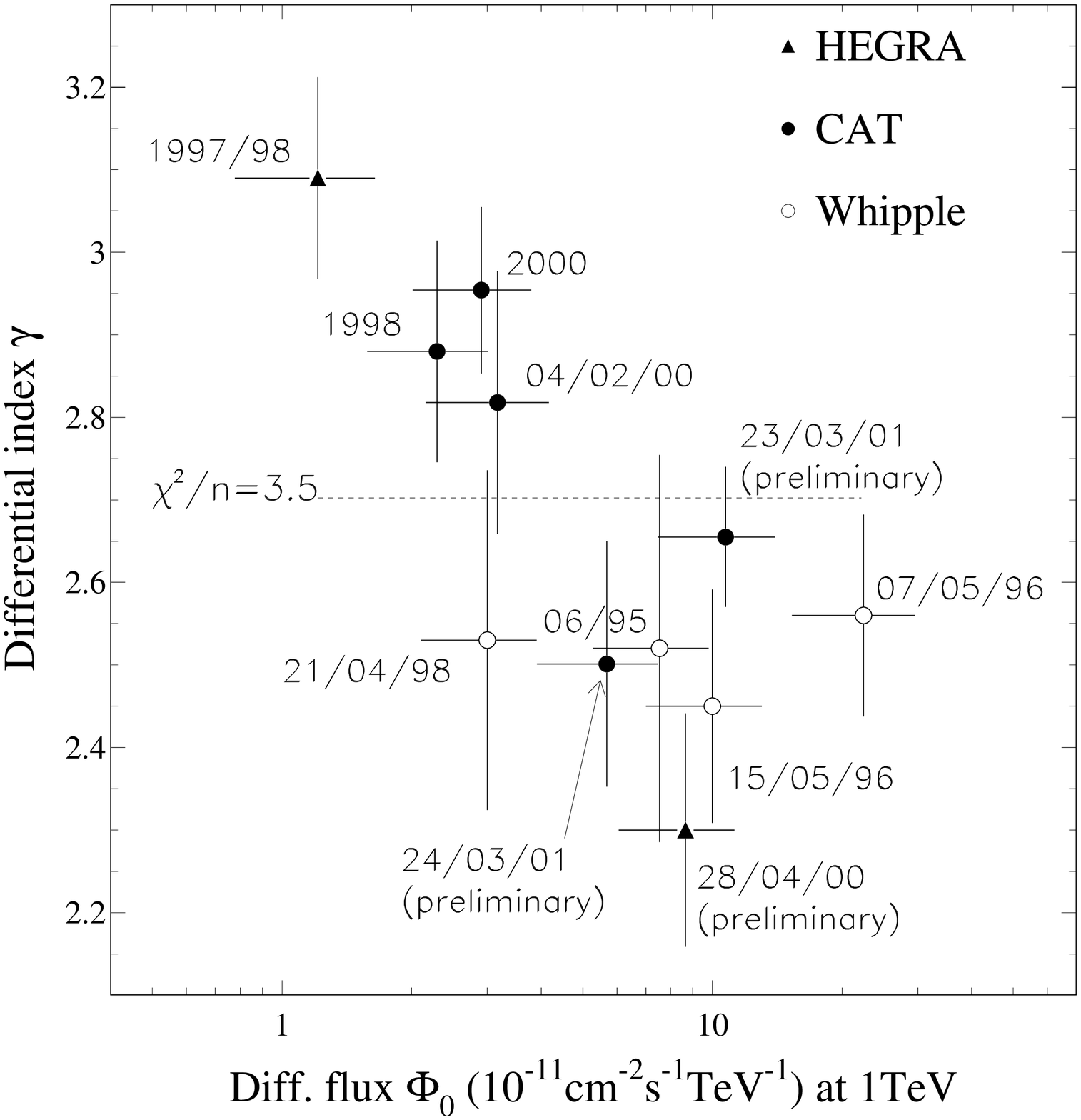,width=0.4\linewidth,clip=
,bbllx=0pt,bblly=0pt,bburx=660pt,bbury=660pt}
{\bf(b)}\hspace*{-.6cm}\epsfig{file=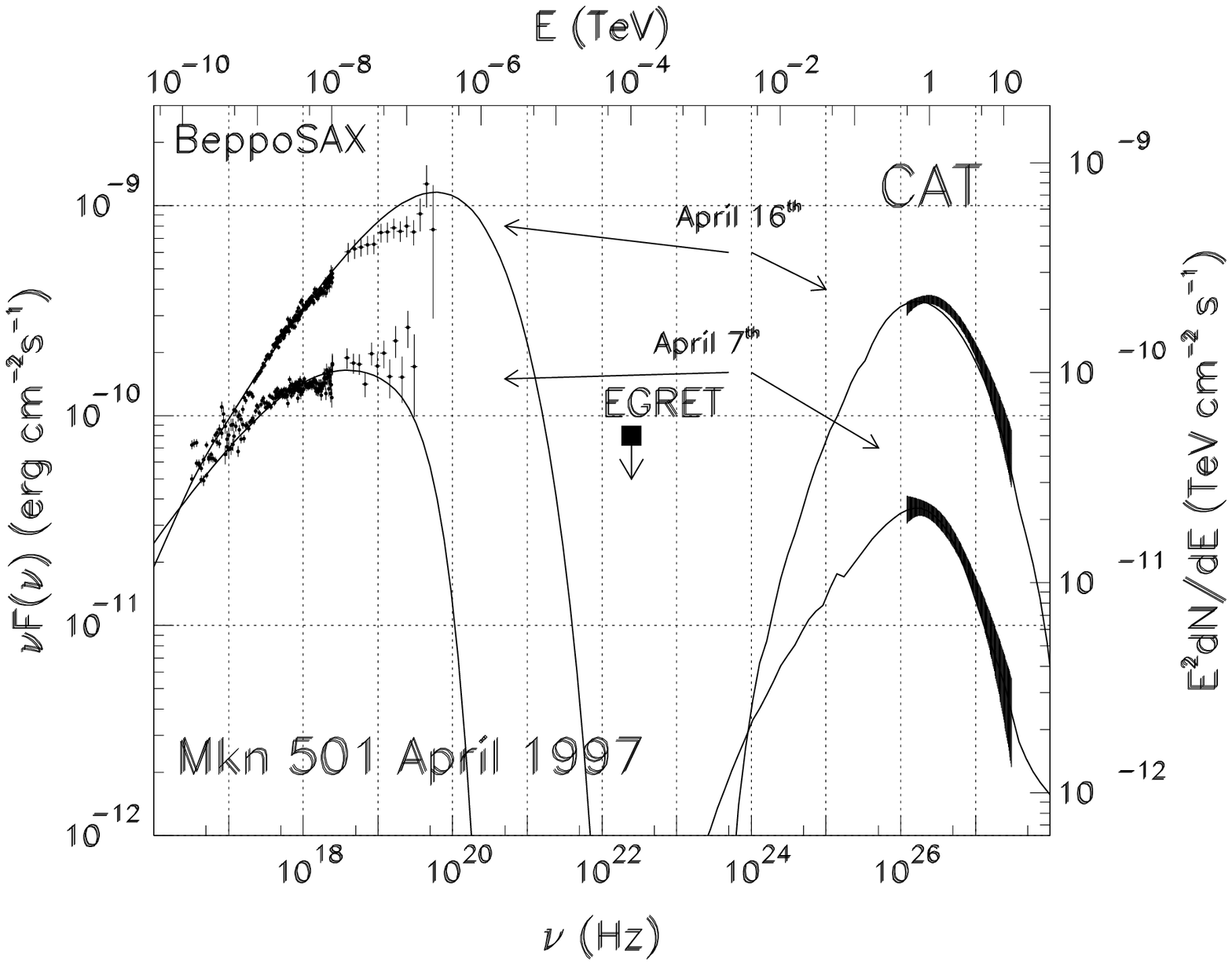,width=0.58\linewidth,clip=
,bbllx=15pt,bblly=0pt,bburx=550pt,bbury=420pt}
\hspace*{0.005\linewidth}
}
\caption[]{
{\bf(a)} Spectral measurements on Mkn~421 by HEGRA, CAT and Whipple between 1995 and 2001.
We added the CAT results obtained during the nights of February 4/5, 2000, and March 24/25, 2001.
The results are given in the $\{\phi_0,\gamma\}$ plane of spectral parameters in the power-law hypothesis.
All error bars take account of both statistical and systematic errors (see~[\cite{Piron01}] for details and references therein).
The $\chi^2/n$ corresponding to the absence of any dependence of the spectral index with flux is $3.5$;
{\bf(b)} Mkn~501 X-ray and VHE SEDs for April 7$^\mathrm{th}$ and 16$^\mathrm{th}$, 1997, as simultaneously measured by
Beppo-SAX and CAT~[\cite{Djannati99,PironThese}]. Full lines come from our homogeneous SSC model~[\cite{PironThese}].
}
\vspace*{-.5cm}
\label{Fig2}
\end{figure}
The light curves of these two blazars, as sampled by CAT during their most active periods, are shown in Fig.~\ref{Fig1}(a).
Both sources are highly variable and a few series of bursts can be distinguished\footnote{The uneven sampling of Mkn~421
light curve in 2001 is due to bad weather conditions.}.
For instance, the flux of Mkn~421 was measured at a level of $(67.3\pm2.6)\times10^{-11}\:\unitphi$
above $250\:\gev$ during the night of March 23/24, 2001 (JD 2451992): this is comparable to the highest $\tev$ flux ever
recorded by CAT, i.e., that of Mkn~501 in the night of April 15/16, 1997 (JD 2450554), which reached
$(83.2\pm3.8)\times10^{-11}\:\unitphi$.
During these flaring periods both sources underwent a large night-to-night variability on several occasions.
Mkn~421 is, however, unique since it has also exhibited important flux variations within a single night.
As an example, its light curves for three nights from the 3$^\mathrm{rd}$ to the 5$^\mathrm{th}$
February 2000 are shown in Fig.~\ref{Fig1}(b). While the fluxes recorded by CAT during the first and last nights were stable, the source
activity changed dramatically in a few hours during the second night: it was reduced by a factor of $2$ in $1$~hour and by a
factor of $5.5$ in $3$~hours. Consequently, the $\gm$-ray emitting zone must be very compact, with a size
$\lesssim$$10$~light-hours if one assumes a typical value of $10$ for the geometric Doppler factor (which reduces the time-scale
in the observer frame). 

As shown in Fig.~\ref{Fig1}(c), Mkn~501 clearly showed a curved
spectrum in 1997, with a $\gm$-ray peak lying above the CAT threshold. Mkn~421 is less extreme: its 1998 time-averaged spectrum
can be well fitted by a simple power-law with a differential index $\gm$$=$$-2.88\pm0.12^\mathrm{stat}\pm0.06^\mathrm{syst}$,
but there is some indication of curvature for the 2000 spectrum due to a larger
signal-to-background ratio~[\cite{Piron01}]. The even higher statistics gathered in the single night of March
23$^\mathrm{rd}$, 2001, yielded a more accurate determination of this curvature, with a differential spectrum
$\displaystyle\frac{\mathrm{d}\phi}{\mathrm{d}E}\varpropto
E_{\mathrm{TeV}}^{-2.77\pm0.09^\mathrm{stat}\pm0.06^\mathrm{syst}-(0.72\pm0.27^\mathrm{stat}\pm0.03^\mathrm{syst})\log_{10}E_{\mathrm{TeV}}}
$.
Still preliminary at the time of the conference, this result has been confirmed since then~[\cite{Khelifi01}].

The question of spectral variability of $\tev$ blazars has been the subject of a long debate. The hardening of the Mkn~501 spectrum
during flares was firstly
discovered by CAT on the basis of a hardness ratio during its 1997 flaring period~[\cite{Djannati99}]. It is illustrated
in Fig.~\ref{Fig1}(c) by the shift in the peak energy of the SED: $0.56$$\pm$$0.10\:\mathrm{TeV}$ for the time-averaged
spectrum and $0.94$$\pm$$0.16\:\mathrm{TeV}$ for the highest flare of April 16$^\mathrm{th}$. Recently, the HEGRA experiment
observed a similar feature by comparing the 1997 and 1998-99 time-averaged spectra~[\cite{Aharonian01}].
To address this question concer\-ning Mkn~421, we compiled its spectral measurements in Fig.~\ref{Fig2}(a), in the $\{\phi_0,\gamma\}$
plane of spectral parameters assuming a power-law shape
$\displaystyle\frac{\mathrm{d}\phi}{\mathrm{d}E}=\phi_0 E_{\mathrm{TeV}}^{-\gm}$
(this hypothesis is justified even in the case of a curved sectrum, see~[\cite{Piron01}] for details).
Despite the general trend seen in this figure, definite conclusions on a possible spectral variability would be
premature: simultaneous observations of more flares by different Cherenkov telescopes are still needed in order to exclude
possible systematic effects between experiments (see~[\cite{Piron01}] for a discussion).

\section{Conclusion}
The multi-wavelength SED of Mkn~501 can be explained by leptonic mo\-dels~[\cite{Ghisellini98}]. The best example to date
is shown in Fig.~\ref{Fig2}(b), where a homogeneous Synchrotron Self-Compton (SSC) model fits well the X-ray and VHE spectra obtained
by the Beppo-SAX satellite and CAT for two dates in April 1997~[\cite{Djannati99}].
In this framework, the CAT observations of Mkn~421 indicate that the peak energy of the inverse Compton contribution to its SED
is significantly lower than the telescope detection threshold.
This is not surprising since {\it (i)} the corresponding synchrotron peak is known to be lower than that of Mkn~501, and
{\it (ii)} leptonic models predict a strong correlation between X-rays and $\gm$-rays. 
This feature is illustrated in Fig.~\ref{Fig2}(b) by the correlated spectral hardening in both energy ranges between
the two dates.
As for Mkn~421, a correlation has been observed many times~[\cite{Maraschi99,Takahashi99}],
but always in terms of integrated (and not differential) fluxes due to the lack of statistics. In addition the VHE spectral variability
of Mkn~421 has not been clearly proven yet.

Moreover, alternative scenarios based on an e$^-$p plasma and proton-induced cascades in the jet are still successful in
interpreting $\tev$ blazars' SEDs~[\cite{Rachen99,Muecke01}].
The study of the dynamic aspects of blazar jet emission, including the temporal and spectral correlations between various
wavelengths, is thus required in the future in order to accurately constrain existing models and to understand the particle
acceleration and cooling processes occurring at the sub-parsec scale in jets. It should help in discriminating between these
models and allow, in particular, to address more deeply the crucial problem of the plasma jet content.

%%-----------------------------
%%      your bibliography
%%-----------------------------

\end{document}